\documentclass{iaus}
\usepackage{graphicx}

\def\spose#1{\hbox to 0pt{#1\hss}}
\def\la{\mathrel{\spose{\lower 3pt\hbox{$\mathchar"218$}}
     \raise 2.0pt\hbox{$\mathchar"13C$}}}
\def\ga{\mathrel{\spose{\lower 3pt\hbox{$\mathchar"218$}}
     \raise 2.0pt\hbox{$\mathchar"13E$}}}
\def\msun{{\rm M}_{\odot}}
\def\rsun{{\rm R}_{\odot}}

\def\citep#1{(\citealt{#1})}  

\def\mearth{{\rm M}_{\oplus}}
\def\rearth{{\rm R}_{\oplus}}

\def\mpers{\rm m\ s^{-1}}

\def\day{\rm day}

\title[The MEarth project] 
{The MEarth project: searching for transiting habitable super-Earths
around nearby M-dwarfs}

\author[Irwin et al.]   
{Jonathan Irwin$^1$, David Charbonneau$^1$, Philip Nutzman$^1$, \\ \and Emilio Falco$^{1}$}

\affiliation{$^1$Harvard-Smithsonian Center for Astrophysics, \\ 60 Garden Street, Cambridge, MA 02138, USA \\ email: {\tt jirwin@cfa.harvard.edu}}

\pubyear{2008}
\volume{253}  
\pagerange{}
\jname{Transiting Planets}
\begin{document}

\maketitle

\begin{abstract}
Due to their small radii, M-dwarfs are very promising targets to
search for transiting super-Earths, with a planet of 2 Earth radii
orbiting an M5 dwarf in the habitable zone giving rise to a 0.5\%
photometric signal, with a period of two weeks.  This can be detected
from the ground using modest-aperture telescopes by targeting samples
of nearby M-dwarfs.  Such planets would be very amenable to follow-up
studies due to the brightness of the parent stars, and the favourable
planet-star flux ratio.  MEarth is such a transit survey of $\sim 2000$
nearby M-dwarfs.  Since the targets are distributed over the entire
(Northern) sky, it is necessary to observe them individually, which
will be done by using 8 independent 0.4m robotic telescopes, two of
which have been in operation since December 2007 at the Fred Lawrence
Whipple Observatory (FLWO) located on Mount Hopkins, Arizona.  We
discuss the survey design and hardware, and report on the current
status of the survey, and preliminary results obtained from the
commissioning data.
\end{abstract}

\firstsection 
\section{Introduction}

M-dwarfs are very favourable targets to search for transiting
super-Earth exoplanets in the habitable zones of their parent stars
(e.g. \cite[Charbonneau \& Deming 2007]{cd07}).  This is predominantly
a result of the very low luminosities of M-dwarfs relative to solar
type stars, which means the habitable zones reside at much smaller
orbital distances.  For the present discussion, we assume an M5-dwarf
star, with mass and radius $\sim 0.25\ \msun$ and $0.25\ \rsun$
respectively, orbited by a planet of $7\ \mearth$ and $2\ \rearth$,
representing a typical target of the MEarth survey.  Given these
parameters, the M-dwarf has a luminosity $\sim 1/200$ that of the Sun,
and consequently, a planet receiving the same stellar insolation as
the Earth would lie at $0.074\ {\rm AU}$, corresponding to an orbital
period of $14.8\ {\rm days}$.  The geometric transit probability for
such a planet is also raised by a factor of approximately three, from
$0.5 \%$ (for the Earth-Sun system) to $1.6\ \%$.

The small radii of M-dwarfs also give rise to substantially deeper
transits ($0.5\ \%$ for the M5-dwarf, compared to $0.03\ \%$ for a
solar-type host), and the combination of a smaller stellar mass and 
shorter orbital period increases the radial velocity semiamplitude,
to $\sim 10\ \mpers$ compared to $1.3\ \mpers$ for the Earth-Sun
system.

The combination of these factors means that the detection of such
planets around M-dwarfs by the transit method is feasible from the
ground, using present-day observational techniques and detectors.
MEarth is such a survey using modest commercially-available equipment,
based on a design study published in \cite{nc08}.

In this contribution, we summarise the results of the design study,
present the novel mode of operation of the MEarth survey, and describe
the implementation and current status of the hardware.

\section{Survey design}

In order to detect a few super-Earth planets, or to place meaningful
constraints on their incidence, for a geometric transit probability of
$\sim 1\%$ we must survey a few thousand M-dwarfs.  These should be
bright in order to allow follow-up studies to be performed, and given
the intrinsic faintness of M-dwarfs this translates to being very
nearby: for example, an M5V star has an absolute J-band magnitude of
$M_J \simeq 9$.  A reasonable limit for follow-up studies is $J \sim
12$, implying distances $< 40\ {\rm pc}$.

Such nearby stars are expected to have high proper motions.  We
therefore appeal to the recently-completed LSPM-North catalogue
(\cite[L\'epine \& Shara 2005]{ls05}), a survey of the entire Northern
hemisphere based on photographic plates, which should be nearly
complete for stars with proper motion $> 0.15''{\rm /yr}$, containing
a total of $\sim 62\,000$ sources.  We further restrict this to a
sub-sample of $4131$ stars within $33\ {\rm pc}$ (\cite[L\'epine
  2005]{l05}) to remove high proper motion sub-dwarf contaminants from
the sample.

In order to select the M-dwarfs from the remaining population of
(predominantly) main sequence dwarfs, we apply colour cuts using
combinations of the $V$-band magnitudes from \cite{ls05}, and
2MASS $J$, $H$ and $K_S$ magnitudes, which are available for every
star in our sample from the 2MASS all-sky data release
(\cite[Skrutskie et al. 2006]{s06}).  The resulting ``culled'' sample
consists of $\sim 3300$ probable M-dwarfs.

These targets are spread in a relatively uniform fashion over the
entire celestial Northern hemisphere.  We have therefore opted for a
strategy of observing them individually.  This mode of operation
brings a number of benefits.  In particular, the requirement on the
field-of-view of the detector is substantially relaxed, since it need
only be large enough to obtain sufficient comparison stars to perform
differential ensemble photometry.  It is also possible to tailor the
parameters of each observation to the individual target M-dwarf, for
example varying the exposure time to achieve the required signal to
noise to detect planets of a given size, thereby saving observing
time.

\cite{nc08} found that the most favourable targets for such a transit
survey are, in fact, the smallest stars: although these are
intrinsically fainter, the reduced count rates are compensated by
having deeper transits, and their faintness increases the number of
suitable comparison stars available for a given field-of-view.  It
is important to recall that for small field-of-view observations of
single targets, the noise in the comparison light curve can become an
important, or even dominant, contributor to the total noise budget.
We therefore further choose to concentrate on the smallest stars,
choosing $1976$ with estimated radii $< 0.33\ \rsun$ (corresponding to
a spectral type of $\sim$ M3).

Since the emission from M-dwarfs peaks in the near-IR, the number of
detected stellar photons would be maximised by observing in this
spectral region.  Unfortunately, near-IR detectors remain
prohibitively expensive, so MEarth uses conventional CCDs.
\cite{nc08} found that the optimal passband for observing the
(extremely red) M-dwarf target stars was a filter cutting on longward
of $\sim 700\ {\rm nm}$, and limited on the red end by the tail of the
CCD quantum efficiency curve.  The H$\alpha$ line was deliberately
omitted from the bandpass due to its potential variability in very
active stars, which is obviously an undesirable feature for precision
photometry.  This passband approximates the sum of transmission of the
Sloan $i$ and $z$ filters (\cite[Fukugita et al. 1996]{f96}), and has
been termed an ``$i+z$'' filter.  In reality, this is formed using a
single piece of Schott RG715 glass, and is therefore economical and
straightforward to manufacture, and the final bandpass using an e2v
CCD42-40 is very similar to the conventional Cousins $I$ filter.

\begin{figure}[h]
\begin{center}
\includegraphics[width=4in]{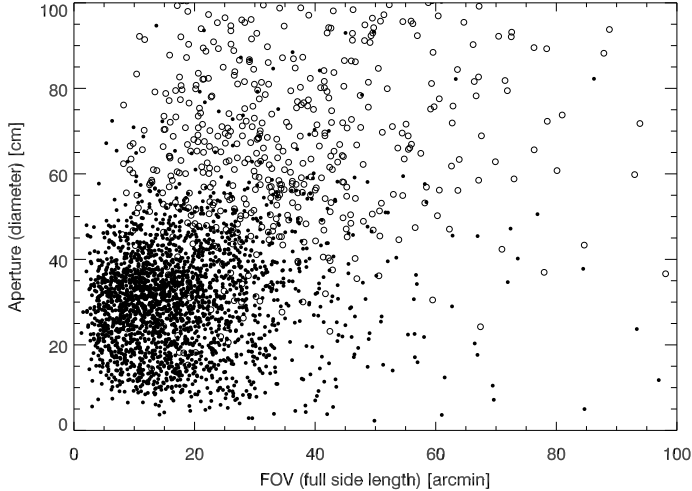}
\caption{Necessary aperture vs. necessary field of view for
  LSPM M-dwarfs.  Stars with radius $>0.33\ \rsun$ are represented by
  open circles, while stars with radius $<0.33\ \rsun$ are represented
  by filled circles.  In order to determine the required telescope
  aperture, we demanded the photometric precision necessary to achieve
  a $3 \sigma$ detection of a transiting 2 $\rearth$ planet in a $150
  s$ integration through the $i+z$ filter.  The range of the x and y
  axes match that of the stars with radii $<0.33\ \rsun$, while $30\%$
  of the M-dwarfs with radii $>0.33\ \rsun$ require more than a $100\
  {\rm cm}$ aperture, and thus fall above the plot limits.}
\label{apfov}
\end{center}
\end{figure}

Given these parameters, we can calculate the telescope aperture and
field-of-view required to survey our target stars.  This was done
assuming a standard noise model, including contributions from Poisson
noise in the stellar counts and sky background, dark current, readout
noise, and atmospheric scintillation using the formulation of
\cite{y67}.  Figure \ref{apfov} indicates that a $40\ {\rm cm}$
aperture telescope with a $25' \times 25'$ field-of-view is sufficient
for the vast majority of the M-dwarfs with $R < 0.33\ \rsun$.

We have opted for eight such telescopes, which leads to an expected
$2\ {\rm yr}$ duration to complete the survey of the habitable zones
of these $1976$ M-dwarfs.  The design study indicates that a yield of
$2.6$ habitable zone super-Earths would be predicted if the true
occurrence of these planets was $10\%$ around our targets, with larger
and closer-in planets being easier to detect.  A null result would
limit the occurrence of $> 2\ \rearth$ super-Earth planets in the
habitable zones of late-M dwarfs to be $< 17\%$ at the $99\%$
confidence level, a result that again becomes a stronger limit for
closer-in planets.

\section{Mode of operation}

\begin{figure}[h]
\begin{center}
\includegraphics[height=5in,angle=270]{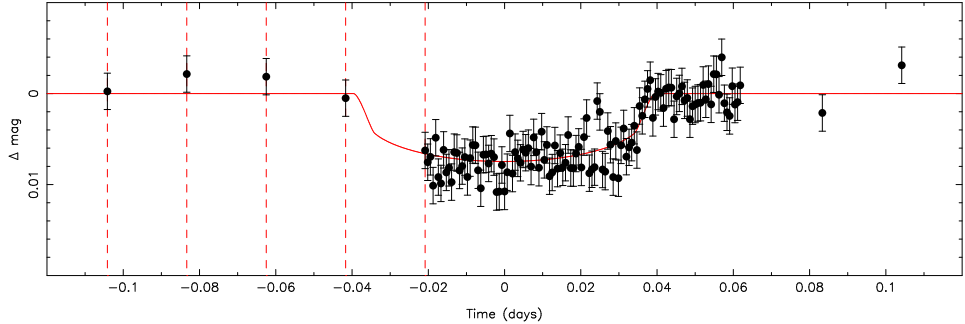}
\caption{Illustration of a possible sampling strategy for transit
  detection with MEarth.  The solid line shows a transit model.  The
  dashed vertical lines indicate the sample times during the normal
  (low-cadence) operation up to the first point which deviates by $> 1
  \sigma$ in the downward direction.  Immediate follow-up is triggered
  at this point, confirming the downward deviation.  Sampling
  continues at high-cadence until after egress, at which point the
  sampling returns to normal until another significant downward light
  curve deviation is detected.}
\label{trdet}
\end{center}
\end{figure}

In order to improve the survey efficiency, and in particular, to
increase the number of M-dwarfs that can be monitored at once by each
telescope, we have adopted a novel detection strategy (illustrated in
Figure \ref{trdet}).  Routine observations will be carried out at
extremely low cadence, with $\sim 2$ visits to each target per transit
timescale (we assume a mid-latitude transit, which leads to a duration
$0.866$ times that of an equatorial transit).  The cadence is further
limited to not being less than once every half-hour to assure a
reasonable number of observations of each target per night, and to
catch shorter-period transits due to planets interior to the habitable
zone.  We intend to detect transits while they are still in progress
via real-time analysis of the images as they are taken.  This
information can then be used to immediately direct follow-up resources
(e.g. other MEarth telescopes) to confirm or reject the hypothesis
that there is a transit in progress.  In practice, this will probably
proceed by immediately obtaining several more data-points, to combat
the effects of noise and light curve systematics, and then following
the remainder of the event, if it turns out to be real, at
high-cadence to obtain a well-sampled transit egress.  By continuously
re-evaluating the probability of there being a transit in-progress
upon obtaining additional data-points, we can quickly reject false
positives without any major effect on the remainder of the targets.
We can therefore cope with a relatively high false alarm rate.

\section{Followup and false positives}

As with any transit survey, in order to confirm the planetary nature
of any detections, radial velocity follow-up observations will be
required.  This should be possible with existing instrumentation, with
several groups demonstrating the required radial velocity precision on
M-dwarfs using conventional instruments in relatively blue passbands.
Ideally, in order to take advantage of the greater continuum flux of
M-dwarfs in the far red and infrared, radial velocities would be
derived from lines in these spectral regions.  Unfortunately the
conventional Iodine cell technique is not usable in the far-red, and
the HARPS (and by extension, HARPS-NEF) spectrographs cannot be used
beyond $\sim 690\ {\rm nm}$.  The new TRES spectrograph on the
Tillinghast $1.5\ {\rm m}$ telescope at FLWO appears promising in this
respect, since it can operate around the atmospheric window close to
$900\ {\rm nm}$, as do several IR radial velocity methods currently
under development, for example \cite[Blake et al. (2007,
  2008)]{bl07,bl08} and the T-EDI project (Lloyd et al., this volume).

It should be noted that the target selection employed for MEarth
largely eliminates the most common sources of astrophysical false
positives that plague conventional wide-field transit surveys.  In
particular, the selection by proper motion eliminates giant host stars
from the sample, and hierarchical triple systems are extremely
unlikely due to the very late spectral types that they must have in
order to be included in the MEarth sample (the brightest star in these
would be a mid-M dwarf due to the colour selection we apply).  Triples
may also be resolvable with high-resolution imaging due to their close
proximity to the Earth. Finally, blends with background binaries are
substantially reduced, not least because our pixel scale of
$0.75''/{\rm pix}$ is $\sim 1/20$ that of a typical wide-field transit
survey, and the high proper motions of our targets mean that previous
epochs of imaging can be used to resolve many of these systems, by
looking for the background source at the present position of the
M-dwarf.

\section{Status}

At the time of writing, MEarth has $5$ telescopes in operation, with
two operating routinely since January 2008, and the final three since
June 2008.  We anticipate having the remaining three telescopes
operational by September 2008.  All of the telescopes are housed in
the same building.  We note that the use of multiple sites at
different longitudes would be beneficial to improve phase coverage,
but would be much more expensive, since we were able to re-use an
existing building and much of the associated support infrastructure at
FLWO, affording a very substantial cost saving.

Data are presently reduced in real-time, although the real-time
analysis software is not yet in-place, using a modified version of the
pipeline from the Monitor project, described in \cite{i07}.  We have
already achieved photometry at the $\sim 0.6\%$ level on M-dwarf
targets over $\sim 90\ {\rm nights}$, and are currently investigating
methods to further reduce the level of systematics present in the
data.  This has been sufficient to detect low-amplitude (few percent
peak-to-peak) rotational modulations in several of our targets for
which sufficient observations are available.

\begin{figure}[h]
\begin{center}
\includegraphics[height=5in,angle=270,bb=44 24 162 747,clip]{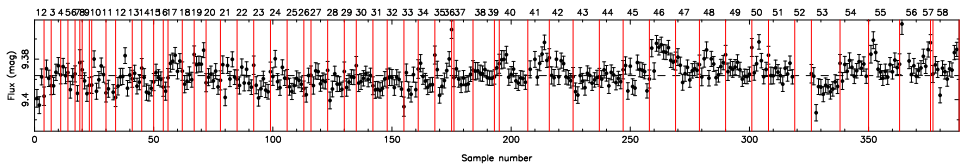}
\caption{Example of an M-dwarf light curve from MEarth data taken
  during the first season of operations, plotted in $I$ magnitude as a
  function of sample number (numbering from 1 for the first data
  point).  Vertical solid lines denote boundaries between different
  nights of observations, which are numbered at the top of the
  diagram.  Data were taken on $58$ separate nights for this object.
  The estimated spectral type of the star is M4V, with an I-band
  magnitude of $\sim 9.4$.}
\label{phot_example}
\end{center}
\end{figure}

\begin{figure}[h]
\begin{center}
\includegraphics[height=5in,angle=270,bb=164 24 362 747,clip]{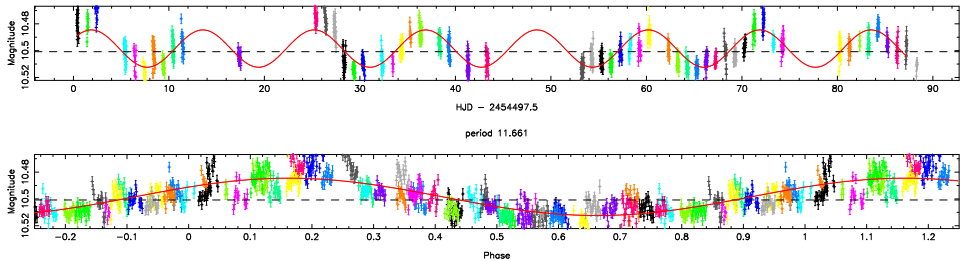}
\caption{Example of an M-dwarf light curve showing periodic
  photometric variations, presumably due to rotation.  The upper panel
  shows the light curve plotted as a function of heliocentric Julian
  day number, and the lower panel shows the curve folded on the
  best-fitting rotation period of $11.661\ {\rm days}$, plotted versus
  orbital phase, where $1$ unit represents a single rotational
  period.  The estimated spectral type of the star is M4.5V, with an
  I-band magnitude of $\sim 10.5$.}
\label{rot_example}
\end{center}
\end{figure}

\begin{figure}[h]
\begin{center}
\includegraphics[height=4in,angle=270]{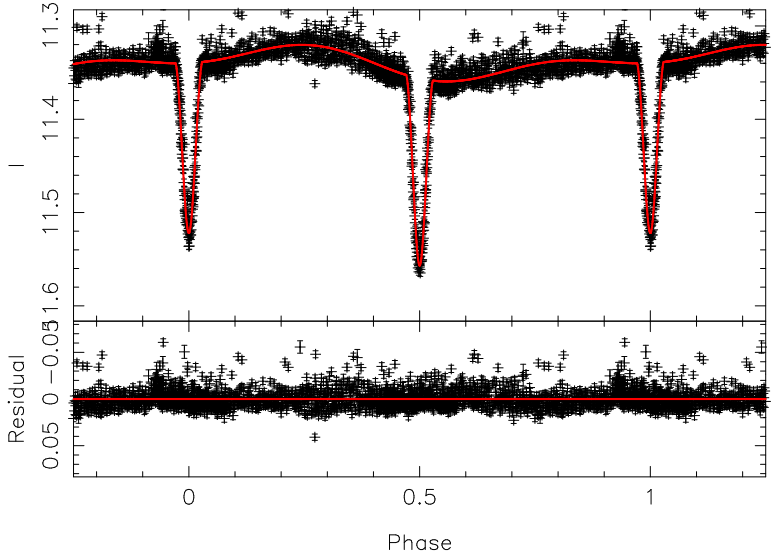}
\caption{Phase-folded light curve of the M-dwarf eclipsing binary
  discovered in MEarth data.  The upper panel shows the $I$-band
  magnitude, over-plotted with an eclipsing binary model (solid line)
  generated using a modified version of {\sc jktebop}
  (\cite[Southworth et al. 2004a,b]{s04a}; \cite[Popper \& Etzel
  1981]{pe81}; \cite[Etzel 1981]{e81}) including out-of-eclipse
  modulations (presumably due to starspots) synchronised with the
  binary orbital period of $0.77\ {\rm days}$. The lower panel shows
  the residuals (data $-$ model) of the fit. Note the differing scales
  on the vertical axes in the two panels.}
\label{eb_lc}
\end{center}
\end{figure}

A new M-dwarf eclipsing binary system was also discovered during the
first few weeks of routine MEarth observations.  The system is a near
equal mass pair, with estimated component masses $\sim 0.3\ \msun$
from photometry, and an extremely short orbital period of $0.77\ {\rm
  days}$.  The light curve of this object obtained using one MEarth
telescope is shown in Figure \ref{eb_lc}.

Such eclipsing binaries are of vital importance for constraining the
mass-radius relation on the main sequence, and hence theoretical
models of low-mass stellar evolution, which are remarkably
poorly-constrained even at old, main sequence ages for stars below $1\
\msun$.  Indeed, many of the observed and well-characterised systems
show substantial discrepancies with the predictions of the theoretical
models (e.g. \cite[Ribas 2006]{rib06}).  These
uncertainties have a direct impact on the reliability 
of stellar parameter estimates, and hence on planet properties
inferred from analysis of transiting systems.  Indeed, as detailed
elsewhere in this volume, many of the transiting planet observations
have reached sufficient quality that this uncertainty in the
parameters of the host star is the limiting factor in our knowledge of
the properties of the planets.  Such M-dwarf eclipsing binaries are
straightforward to detect from MEarth data, and it should be possible
to characterise them extremely well due to their brightness, allowing
model-independent mass and radius estimates with uncertainties of
only a few percent to be made (e.g. \cite[Andersen 1991]{a91}).

\section*{Acknowledgments}

It is a pleasure to acknowledge the assistance of the staff at the
Fred Lawrence Whipple Observatory in Arizona: Karen Erdman-Myres,
Grace Alegria, Rodger Harris, Dave Martina, Dennis Jankovsky, Tom
Welsh, Wayne Peters, Ted Groner, Bob Hutchins, Perry Berlind, Mike
Calkins, and Gil Esquerdo; and the Harvard-Smithsonian Center for
Astrophysics: Irene Coyle, Andrea Moore, Maureen Connors, Jean
Collins, Sara Yorke, and Leslie Feldman.  We acknowledge funding from
the David and Lucile Packard Fellowship for Science and Engineering,
Harvard University, and the U.S. National Science Foundation.  This
work has made use of data products from the Two Micron All Sky Survey,
which is a joint project of the University of Massachusetts and the
Infrared Processing and Analysis Center / California Institute of
Technology, funded by the National Aeronautics and Space
Administration and the National Science Foundation.

\end{document}